\documentclass[aps,twocolumn,showpacs,superscriptaddress]{revtex4}

\usepackage{graphicx}
\usepackage{bm}
\usepackage{amsmath}
\usepackage{dcolumn}%

\newcommand{\eg}{\ensuremath{e_{g}}}
\newcommand{\tg}{\ensuremath{t_{2g}}}

\newcommand{\mb}{\ensuremath{\mu_{\text{B}}}}

\begin{document}

\title{Electronic structure and x-ray magnetic circular dichroism in the
  quadruple perovskite CaCu$_3$Re$_2$Fe$_2$O$_{12}$ }

\author{ L.V. Bekenov}

\affiliation{G. V. Kurdyumov Institute for Metal Physics of the
  N.A.S. of Ukraine, 36 Academician Vernadsky Boulevard, UA-03142
  Kyiv, Ukraine}

\author{ D.V. Mazur}

\affiliation{G. V. Kurdyumov Institute for Metal Physics of the
  N.A.S. of Ukraine, 36 Academician Vernadsky Boulevard, UA-03142
  Kyiv, Ukraine}

\author{B.F. Zhuravlev}

\affiliation{G. V. Kurdyumov Institute for Metal Physics of the
  N.A.S. of Ukraine, 36 Academician Vernadsky Boulevard, UA-03142
  Kyiv, Ukraine}

\author{S.V. Moklyak}

\affiliation{G. V. Kurdyumov Institute for Metal Physics of the
  N.A.S. of Ukraine, 36 Academician Vernadsky Boulevard, UA-03142
  Kyiv, Ukraine}

\author{Yu.N. Kucherenko}

\affiliation{G. V. Kurdyumov Institute for Metal Physics of the
  N.A.S. of Ukraine, 36 Academician Vernadsky Boulevard, UA-03142
  Kyiv, Ukraine}

\author{V.N. Antonov}

\affiliation{G. V. Kurdyumov Institute for Metal Physics of the
  N.A.S. of Ukraine, 36 Academician Vernadsky Boulevard, UA-03142
  Kyiv, Ukraine}

\date{\today}

\begin{abstract}

We have studied the electronic and magnetic properties of the A- and B-site-ordered 
perovskite CaCu$_3$Re$_2$Fe$_2$O$_{12}$ within the density-functional theory
using the generalized gradient approximation (GGA) with the consideration of
strong Coulomb correlations (GGA+$U$) in the framework of the fully
relativistic spin-polarized Dirac linear muffin-tin orbital band structure
method. We have calculated the x-ray absorption spectroscopy (XAS) and x-ray
magnetic circular dichroism (XMCD) spectra at the Cu, Fe, Re $L_{2,3}$ and
O $K$ edges. The calculated results are in good agreement with experiment.
We show that the GGA+$U$ method produces better agreement with the
experimental spectra if Hubbard $U$ is applied to Cu and Fe sites.

\end{abstract}


\maketitle

\section{Introduction}

\label{sec:introd}

Transition-metal perovskites have been studied for half a century, and most
intensively during the last decade, for their fascinating electronic and
magnetic properties arising from narrow 3$d$ bands and strong Coulomb
correlations \cite{JoSa50,WoCo50,Good55,MoIm04,IFT98}. Perovskite-structure
oxides with the general formula ABO$_3$ display a large variety of intriguing
properties and raise lots of important fundamental issues in solid-state
physics and chemistry, such as high-temperature superconductivity
\cite{BeMu86}, colossal magnetoresistivity \cite{JoSa50,SaJo50},
ferroelectricity \cite{Bus12}, and multiferroicity \cite{PSM05}. The structure
can be described as a network of corner-sharing BO$_6$ octahedra, the oxygen ions
of which provide a twelve-coordinated environment for large A ions. 
The physical properties of perovskite oxides can be tuned by chemical
substitution. For example, the double perovskite structure
A$_2$BB$^{\prime}$O$_6$ with two different kinds of B-site transition-metal
ions demonstrates broad compositional diversity \cite{STI07}.

A-site ordered perovskites with the general formula AA$^{\prime}_3$B$_4$O$_{12}$, 
called sometimes quadruple perovskites \cite{TYO13}, are a derivative of perovskites. 
Such perovskite-structure oxides can be obtained by 
filling 3/4 of A sites with small transition-metal cations and the other 1/4 with 
larger alkali, alkaline earth, or rare earth cations. They have a 2$a \times 2a \times 2a$ 
crystal structure with A$^{\prime}$O$_4$ square-planar units stabilized by heavily tilted BO$_6$ octahedra. 
Introduction of A$^{\prime}$ - A$^{\prime}$ and A$^{\prime}$ - B magnetic interactions 
in addition to B-B coupling gives rise to a variety of intriguing properties
\cite{Shim08,CMS+14}. Cu and Mn are typical transition-metal cations readily
accommodated by the A$^{\prime}$ site. Compounds with this structure type were
synthesized in the 1960s and 1970s \cite{DRT67,MDC+73}, and recently lots of
fascinating functional properties have been discovered in them. Among those properties are 
a large dielectric constant in CaCu$_3$Ti$_4$O$_{12}$ \cite{HVS+01}, a large negative 
thermal-expansion-like volume change due to intersite charge transfer in LaCu$_3$Fe$_4$O$_{12}$
\cite{LHS+09}, and multiferroism in CaMn$_3$Mn$_4$O$_{12}$ \cite{JCK+12}. In
CaCu$_3$B$_4$O$_{12}$ materials with nonmagnetic ions at the B site, the
A$^{\prime}$ - A$^{\prime}$ interaction is ferromagnetic for B = Ge and Sn but
antiferromagnetic for B = Ti \cite{SSY+07,SSS08}. When magnetic species like
Mn are introduced at the B sites, the A$^{\prime}$(Cu)-B(Mn)
antiferromagnetic interaction becomes dominant and ferrimagnetism is observed
in ACu$_3$Mn$_4$O$_{12}$ for A = Ca, La and Bi \cite{ZGS+99,ASD+03,TYA+07}.

More complex A- and B-site-ordered quadruple perovskite-structure oxides
with the general formula AA$^{\prime}_3$B$_2$B$^{\prime}_2$O$_{12}$ (referred
to as 1322-type) display a large variety of intriguing properties. These
materials have the cubic structure (space group $Pn\bar{3}$) with 1:3 order of
12-coordinated A and 4-coordinated square-planar A$^{\prime}$ cations, and are
formed predominantly for A$^{\prime}$ = Cu. Ferrimagnets with Curie
temperatures $T_C$ $>$ 300 K are found among CaCu$_3$B$_2$B$^{\prime}_2$O$_{12}$ 
(for B/B$^{\prime}$ = Fe/Re, Fe/Os, Mn/Os) \cite{RSG+00,HVS+01,LHS+09} and 
ACu$_3$Fe$_2$Os$_2$O$_{12}$ (for A = Na, La) \cite{ShTa09,CSH+12} materials, 
while ferri- or antiferromagnetic order at lower temperature is observed in other analogies, 
such as CaCu$_3$B$_2$B$^{\prime}_2$O$_{12}$ with B/B$^{\prime}$ = 
Ga/Sb, Cr/Sb, Fe/Sb, Fe/Nb \cite{ZDY+11,JCK+12,Shi08}, LaCu$_3$Co$_2$Re$_2$O$_{12}$ \cite{Roy54} and
LaMn$_3$Ni$_2$Mn$_2$O$_{12}$ \cite{HKW59}. The latter is the only A$^{\prime}$
= Mn (and non-Cu) 1322 phase reported to date \cite{SKR+22}. 
CaCu$_3$Cr$_2$Sb$_2$O$_{12}$ and CaCu$_3$Fe$_2$Sb$_2$O$_{12}$ have nonmagnetic
Sb$^{5+}$ ions at the B$^{\prime}$ site so antiferromagnetic
A$^{\prime}$(Cu$^{2+}$)-B(Cr$^{3+}$/Fe$^{3+}$) interactions are dominant and
these materials are ferrimagnetic insulators with $T_c$ of 160 and 170 K,
respectively \cite{BLP+05,CMS+13}. In contrast, CaCu$_3$Cr$_2$Ru$_2$O$_{12}$
is metallic but shows Pauli-paramagnetic behavior \cite{BLP+05}.

In the paper, we carry out a detailed {\it ab initio} study of the electronic structure, XAS and XMCD
spectra of the quadruple perovskite CaCu$_3$Fe$_2$Re$_2$O$_{12}$ (CCFRO) in
terms of the density functional theory. Our study sheds light on the important
role of band structure effects and transition metal $d$ $-$ oxygen 2$p$
hybridization in the spectral properties of this transition metal oxide. We use
the fully relativistic spin-polarized Dirac linear muffin-tin orbital band structure method. 
Both the generalized gradient approximation (GGA) and the GGA+$U$ approach are employed to evaluate the
sensitivity of the XAS and XMCD results to different treatment of
correlated electrons.

The experimental measurements of the XAS and XMCD properties of CCFRO
were provided by Shimakawa and Mizumak \cite{ShMi14}. The
authors obtained the XAS and XMCD spectra at the Cu, Fe, and Re $L_{2,3}$
edges, as well as estimated the values of magnetic moments by using the so called sum rules for XMCD spectra. 
The observed saturated magnetization of 8.7 {\mb}/f.u is close to a collinear
ferrimagnetic Ca$^{2+}$Cu$^{2+}_3$($\uparrow$)Fe$^{3+}_2$($\uparrow$)Re$^{5+}_2$($\downarrow$)O$_{12}$
value of 9 {\mb}/f.u., neglecting orbital contributions. This ferrimagnetic
spin structure is confirmed by the XMCD and XAS measurements.

The paper is organized as follows. The computational details are presented in
Sec. II. Section III presents the electronic and magnetic structures of
CaCu$_3$Fe$_2$Re$_2$O$_{12}$. Section IV presents the theoretically calculated
XAS and XMCD spectra at the Cu, Fe, Re $L_{2,3}$ and oxygen $K$ edges. The theoretical results are
compared with experimental measurements. Finally, the results are summarized
in Sec. V.

\section{Crystal structure and computational details}
\label{sec:details}

\subsection{X-ray magnetic circular dichroism} 

Magneto-optical (MO) effects refer to various changes in the
polarization state of light upon interaction with materials possessing
a net magnetic moment, including rotation of the plane of linearly
polarized light (Faraday, Kerr rotation), and the complementary
differential absorption of left and right circularly polarized light
(circular dichroism). In the near visible spectral range these effects
result from excitation of electrons in the conduction band. Near x-ray
absorption edges, or resonances, the magneto-optical effects can be
enhanced by transitions from well-defined atomic core levels to empty
valence or conduction states.

Within the one-particle approximation, the absorption coefficient
$\mu^{\lambda}_j (\omega)$ for incident x-ray polarization $\lambda$ and
photon energy $\hbar \omega$ can be determined as the probability of
electronic transitions from initial core states with the total angular
momentum $j$ to final unoccupied Bloch states

\begin{eqnarray}
\mu_j^{\lambda} (\omega) &=& \sum_{m_j} \sum_{n \bf k} | \langle \Psi_{n \bf k} |
\Pi _{\lambda} | \Psi_{jm_j} \rangle |^2 \delta (E _{n \bf k} - E_{jm_j} -
\hbar \omega ) \nonumber \\
&&\times \theta (E _{n \bf k} - E_{F} ) \, ,
\label{mu}
\end{eqnarray}
where $\Psi _{jm_j}$ and $E _{jm_j}$ are the wave function and the
energy of a core state with the projection of the total angular
momentum $m_j$; $\Psi_{n\bf k}$ and $E _{n \bf k}$ are the wave
function and the energy of a valence state in the $n$-th band with the
wave vector {\bf k}; $E_{F}$ is the Fermi energy.

$\Pi _{\lambda}$ is the electron-photon interaction
operator in the dipole approximation
\begin{equation}
\Pi _{\lambda} = -e \mbox{\boldmath$\alpha $} \bf {a_{\lambda}},
\label{Pi}
\end{equation}
where $\bm{\alpha}$ are the Dirac matrices and $\bf {a_{\lambda}}$ is the
$\lambda$ polarization unit vector of the photon vector potential, with
$a_{\pm} = 1/\sqrt{2} (1, \pm i, 0)$, $a_{\parallel}=(0,0,1)$. Here, $+$ and
$-$ denotes, respectively, left and right circular photon polarizations with
respect to the magnetization direction in the solid. Then, x-ray magnetic
circular and linear dichroism are given by $\mu_{+}-\mu_{-}$ and
$\mu_{\parallel}-(\mu_{+}+\mu_{-})/2$, respectively.  More detailed
expressions of the matrix elements in the electric dipole approximation may be
found in Refs.~\cite{GET+94,book:AHY04,AHS+04}. The matrix elements due to
magnetic dipole and electric quadrupole corrections are presented in
Ref.~\cite{AHS+04}.

In order to simplify the comparison of the theoretical x-ray isotropic
absorption spectra of CaCu$_3$Fe$_2$Re$_2$O$_{12}$ with the experimental ones, we
take into account the background intensity, which affects the high energy part
of the spectra and is caused by different kinds of inelastic scattering of the
electron promoted to the conduction band due to the x-ray
absorption (the scattering on potentials of surrounding atoms, defects, phonons
etc.). To calculate the background spectra we used the model proposed by
Richtmyer~{\it et~al.} \cite{RBR34}. The absorption coefficient for the
background intensity is

\begin{equation}
\mu(\omega) = \frac{C \Gamma_c}{2\pi}\int_{E_{cf_0}}^{\infty}
\frac{dE_{cf}}{(\Gamma_c/2)^2+(\hbar \omega-E_{cf})^2}~
\label{mu_Bgr}
\end{equation}
where $E_{cf}=E_c - E_f$, $E_c$, and $\Gamma_c$ are the energy and 
lifetime broadening of the core hole, $E_f$ is the energy of the final empty
continuum level, $E_{cf_0}$ is the energy of the lowest unoccupied
continuum level, and $C$ is a normalization constant, which was used as 
an adjustable parameter in this paper.

Concurrent with the development of the x-ray magnetic circular
dichroism technique, some important magneto-optical sum rules were
derived \cite{LT88,TCS+92,CTA+93,LT96}.

For the $L_{2,3}$ edges the sum rule for the expectation value of the $z$ component of the orbital
angular momentum $\langle l_z\rangle$ can be written as \cite{book:AHY04}

\begin{equation}
\langle l_z\rangle = n_h ~ \frac{4}{3} ~ \frac{\int_{L_3 + L_2} d \omega(\mu_+ - \mu_-)} {
\int_{L_3 + L_2} d \omega(\mu_+ + \mu_-)} ~
\label{l_z}
\end{equation}
where $n_h$ is the number of holes in the $d$ band $n_h=10 - n_d$. The integration is taken over the whole 2$p$ absorption
region. The $s_z$ sum rule can be written as

\begin{eqnarray}
&&\langle s_z\rangle +\frac{7}{2}\langle t_z\rangle =
\nonumber \\  
&& n_h ~ \frac{ \int_{L_3} d \omega(\mu_{+} - \mu_{-}) -2 \int_{L_2} d
\omega(\mu_{+} - \mu_{-})} {\int_{L_3 + L_2} d \omega(\mu_{+} + \mu_{-})}~
\label{s_z}
\end{eqnarray}
\noindent
where $t_z$ is the $z$ component of the magnetic dipole operator ${\bf
t}= {\bf s} - 3 ~ {\bf r} ~ ({\bf r}\cdot {\bf s})/|{\bf r}|^2$, which
accounts for the asphericity of the spin moment. The integration
$\int_{L_3}$ $\left ( \int_{L_2}\right )$ is taken only over the 2$p_{3/2}$
$\left( 2p_{1/2}\right)$ absorption region.

\subsection{Crystal structure} 

\begin{figure}[tbp!]
\begin{center}
\includegraphics[width=0.98\columnwidth]{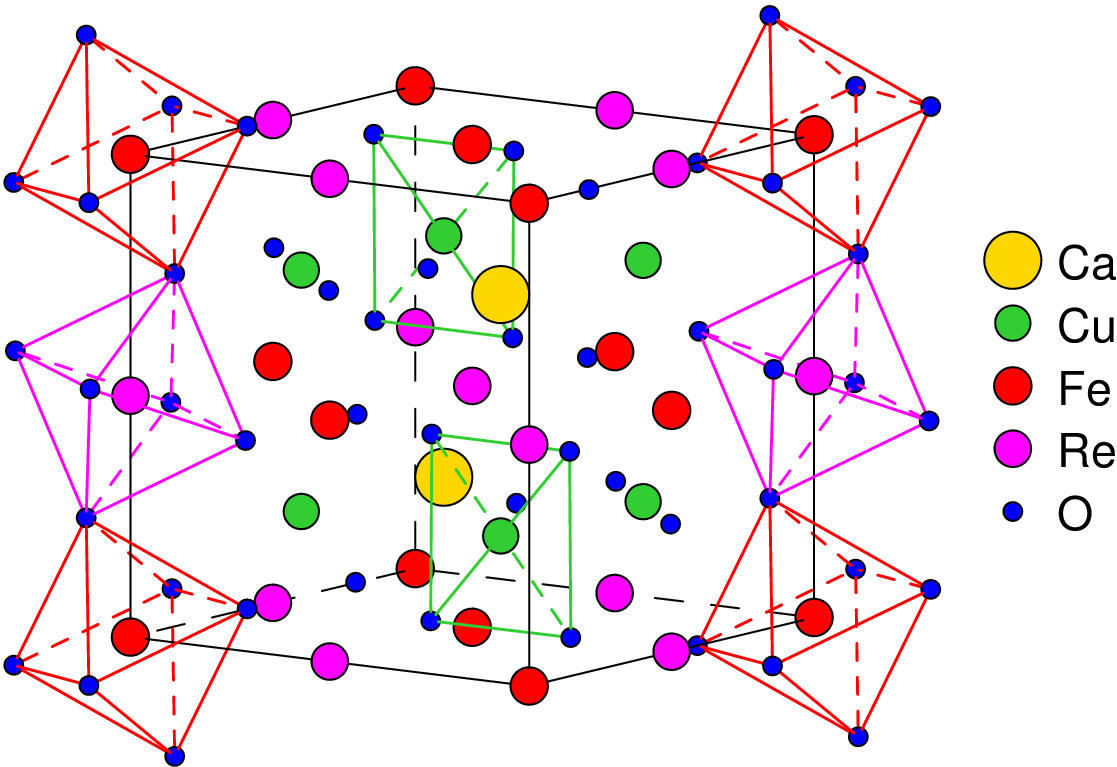}
\end{center}
\caption{\label{struc_CCFRO}(Color online) The crystal structure of the A- and
  B-site-ordered quadruple perovskite CaCu$_3$Fe$_2$Re$_2$O$_{12}$ (the space
  group is $Pn\bar{3}$, No. 201). Ca and Cu ions are ordered at
  the A sites and Fe and Re ions are ordered in a rock-salt-type arrangement
  at the B sites of the ABO$_3$ perovskite structure, resulting in a network
  of CuO$_4$ square units and heavily tilted FeO$_6$ and ReO$_6$ octahedra. }
\end{figure}

The crystal structure of CaCu$_3$Fe$_2$Re$_2$O$_{12}$ (Fig. \ref{struc_CCFRO})
is a variant of the cubic perovskite oxide ABO$_3$. The
superstructure AA$^{\prime}_3$B$_2$B$^{\prime}_2$O$_{12}$ with space group
$Pn\bar{3}$ (No. 201) is formed by quadrupling the parent unit cell and
replacing 3/4 of the element A with A$^{\prime}$. Unlike the A site in the
basic ABO$_3$ perovskite, which is usually occupied by alkali-metal,
alkaline-earth or rare-earth cations, the A$^{\prime}$ site in
AA$^{\prime}_3$B$_2$B$^{\prime}_2$O$_{12}$ can accommodate transition metal
ions. Due to the introduction of A$^{\prime}$, the symmetry of the structure
is lowered by a large rotation of BO$_6$ and B$^{\prime}$O$_6$ octahedra, which
brings four oxygen ions closer to the A$^{\prime}$ site to form a
seemingly nearly square-planar environment.

The cation valences in CaCu$_3$Fe$_2$Re$_2$O$_{12}$, estimated using the observed 
cation-oxygen bond distances from Ref. \cite{CMS+14} and a bond valence sum method, 
were found to be 2.10, 1.90, 3.12, and 4.91 for Ca, Cu, Fe, and Re, respectively \cite{BrAl85}. 
The bond valence sum values are very close to the formal values in the formula
Ca$^{2+}$Cu$^{2+}_3$Fe$^{3+}_2$Re$^{5+}_2$O$_{12}$. The charge difference
between Fe$^{3+}$ and Re$^{5+}$ results in the high degree of rock-salt-type
cation ordering at the B/B$^{\prime}$ sites. The similar formal charge
distribution was observed in CaCu$_3$Fe$_2$Sb$_2$O$_{12}$ \cite{CMS+13} and in
the low-temperature charge-ordered phase of CaCu$_3$Fe$_4$O$_{12}$, where
B-site charge disproportionation below 210 K stabilizes the
Ca$^{2+}$Cu$^{2+}_3$Fe$^{3+}_2$Fe$^{5+}_2$O$_{12}$ configuration
\cite{YTH+08}. We should mention that small inversion disorder of 6.2\% was
found in CaCu$_3$Fe$_2$Re$_2$O$_{12}$ \cite{CMS+14}.

\begin{table}[tbp!]
  \caption {Atomic positions of CaCu$_3$Fe$_2$Re$_2$O$_{12}$ (the lattice constant
    $a$ = 7.44664 \AA \cite{CMS+14}) and metal-oxygen distances. }
\label{struc_tab_CCFRO}
\begin{center}
\begin{tabular}{|c|c|c|c|c|c|c|}
\hline
Atom &Site   & $x$    & $y$    & $z$    & M$-$O (\AA) \\
\hline
 Ca  & 2$a$  & 0.25   & 0.25   & 0.25   & 3.26370 $\times$ 12 \\
 Cu  & 6$d$  & 0.25   & 0.75   & 0.75   & 2.00629 $\times$ 4 \\
 Fe  & 4$b$  & 0      & 0      & 0      & 2.00144 $\times$ 6 \\
 Re  & 4$c$  & 0.5    & 0.5    & 0.5    & 1.93381 $\times$ 6 \\
 O   & 24$h$ & 0.4496 & 0.7548 & 0.0691 &  $-$      \\
\hline
\end{tabular}
\end{center}
\end{table}

\subsection{Calculation details}

The details of the computational method are described in our previous
papers \cite{LYA+06,AHY+07b,AYJ10,RTR+11} and here we only mention
several aspects. The band structure calculations were performed using
the fully relativistic linear muffin-tin orbital (LMTO) method
\cite{And75,book:AHY04}, which uses four-component basis functions
constructed by solving the Dirac equation inside an atomic sphere
\cite{NKA+83}. The exchange-correlation functional of a GGA-type was
used in the version of Perdew, Burke and Ernzerhof (PBE)
\cite{PBE96}. The Brillouin zone (BZ) integration was performed using
the improved tetrahedron method \cite{BJA94} and the self-consistent
charge density was obtained with 1728 {\bf k}-points in the BZ. The
basis consisted of Ca, Cu, Fe, and Re $s$, $p$, $d$, and $f$ and O $s$,
$p$, and $d$ LMTO's.

We found that the agreement between the theoretically calculated and
experimentally measured XAS and XMCD spectra becomes much better with taking
into account strong Coulomb correlations. To include them
into the consideration we used the "relativistic" generalization
of the rotationally invariant version of the LSDA+$U$ method \cite{YAF03}
which takes into account the spin-orbit coupling so that the occupation
matrix of localized electrons becomes non-diagonal in spin indexes. This
method is described in detail in our previous paper \cite{YAF03} including the
procedure to calculate the screened Coulomb $U$ and exchange $J$ integrals, as
well as the Slater integrals $F^2$, $F^4$, and $F^6$. In our calculations, the
intrashell Coulomb repulsion $U$ and interorbital Hund's magnetic coupling $J$
were applied to Cu, Fe, and Re sites: $U_{Cu}$ = 6 eV, $J_{Cu}$ = 1.0 eV,
$U_{Fe}$ = 4 eV, $J_{Fe}$ = 0.9 eV, $J_{Re}$ = 2 eV, $J_{Re}$ = 0.7 eV. Thus,
the parameter $U_{\rm{eff}} = U - J$ which roughly determines the splitting
between the lower and upper Hubbard bands, was $U^{Cu}_{\rm{eff}}$ = 5,
$U^{Fe}_{\rm{eff}}$ = 3.1, and $U^{Re}_{\rm{eff}}$ = 1.3 eV.

The x-ray absorption and dichroism spectra were calculated taking into
account the exchange splitting of core levels.
The finite lifetime of a core hole was accounted for by folding the spectra
with a Lorentzian. The widths $\Gamma$ of $L_{2,3}$ core levels for Cu, Fe,
and Re were taken from Ref. \cite{CaPa01}. The finite experimental resolution
of the spectrometer was accounted for by a Gaussian of 0.6 eV (the $s$
coefficient of the Gaussian function).

\section{Electronic structure}

\begin{figure}[tbp!]
\begin{center}
\includegraphics[width=0.95\columnwidth]{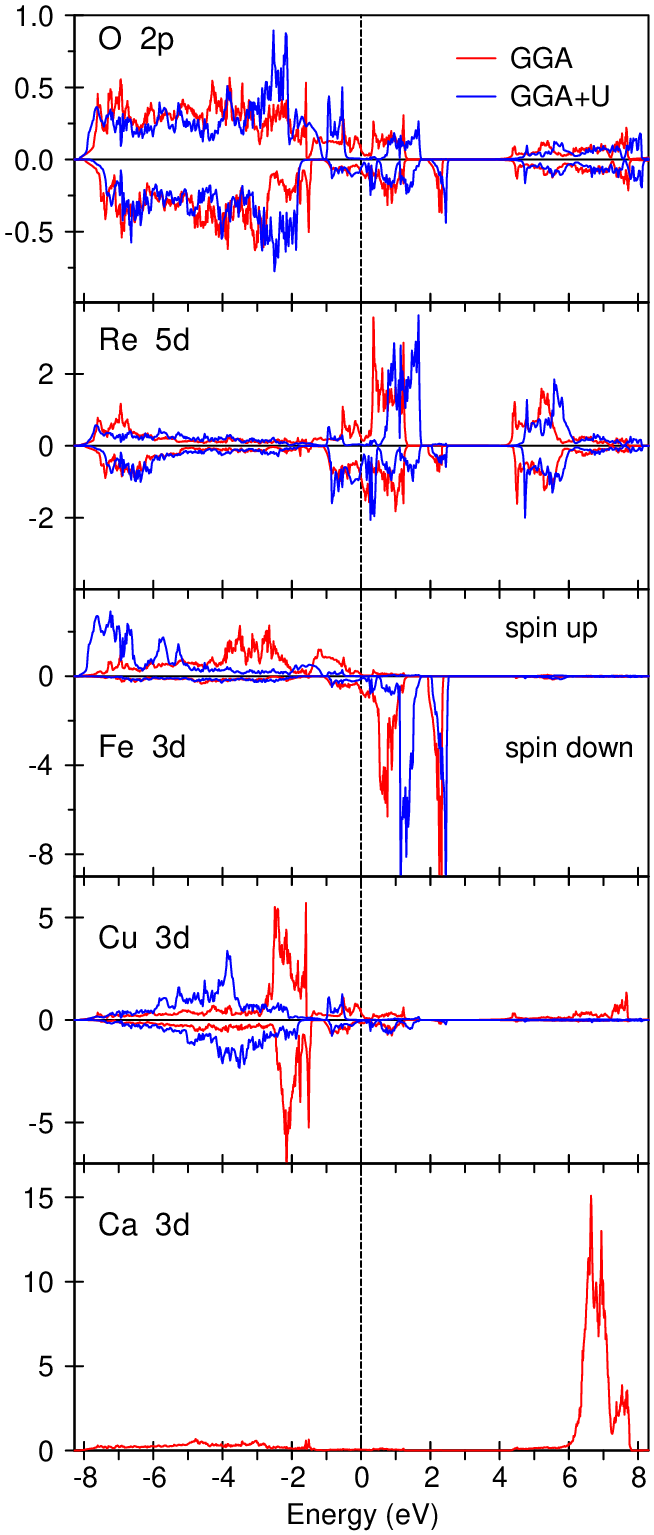}
\end{center}
\caption{\label{PDOS_CCFRO}(Color online) The partial density of states [in
    states/(atom eV)] of CaCu$_3$Fe$_2$Re$_2$O$_{12}$ calculated in the GGA
  (red curves) and GGA+$U$ (blue curves) approximations (for
  $U^{Cu}_{\rm{eff}}$ = 5, $U^{Fe}_{\rm{eff}}$ = 3.1, and $U^{Re}_{\rm{eff}}$
  = 1.3 eV). }
\end{figure}

Figure \ref{PDOS_CCFRO} presents the partial density of states for
CaCu$_3$Fe$_2$Re$_2$O$_{12}$ calculated in the GGA and GGA+$U$
approximations. The Cu 3$d$ valence states are situated from $-$8.1 to 1.8
eV and from 4.1 to 7.7 eV. The Fe spin-down states are mostly empty, and the spin-up
states occupy the energy interval from $-$8.9 up to 1.1 eV. The empty Fe spin-down
states possess two peaks from $E_F$ to 1.2 eV and from 1 to 2.3
eV (in GGA). The Re 5$d$ states consist from 
an almost similar amount of spin-up and spin-down states. The Ca 3$d$ empty
states occupy the energy interval 6.1$-$7.8 eV. The O 2$s$ states are located
mostly between $-$19.7 and $-$17.4 eV below the Fermi energy (not shown). The
O 2$p$ states are situated from $-$8.2 to 2.8 eV and from 4.5 to 8.2 eV. They
are well hybridized with Re 5$d$ and Fe 3$d$ states.  The spin splitting of
the oxygen 2$p$ states is quite small (around 0.2 eV).

In the GGA+$U$ approach (the blue curves in Fig. \ref{PDOS_CCFRO}), Hubbard $U$ shifts 
the occupied and empty states by $U_{\rm{eff}}$/2 downwards and upwards, respectively. 
As a result, energy gaps appear in the spin-up states at the Cu, Fe, Re, and oxygen sites. 
Besides, the Cu and Fe 3$d$ states spread in a wider energy interval and change the shape. 
Although Hubbard $U$ was not applied to oxygen states, they are also changed due to
the strong hybridization with Fe 3$d$ and Re 5$d$ states. We can expect a strong effect of 
Hubbard $U$ on the XAS and XMCD spectra (see the next section).
 
\section{X-ray absorption and XMCD spectra}
\subsection{Cu, Fe, Re $L_{2,3}$ and O $K$ spectra}

\begin{figure}[tbp!]
\begin{center}
\includegraphics[width=0.95\columnwidth]{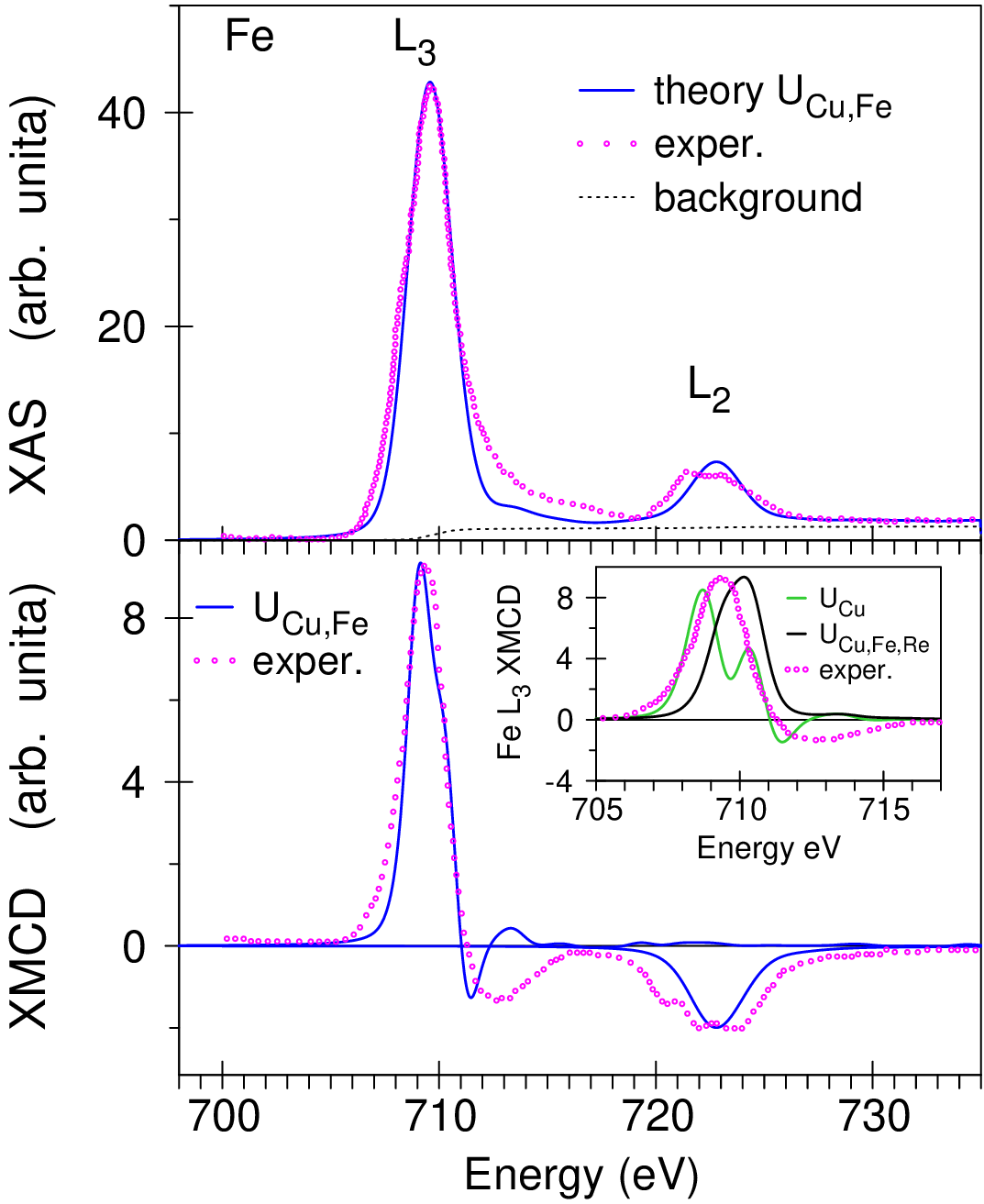}
\end{center}
\caption{\label{xmcd_Fe_CCFRO}(Color online) Upper panel: the experimental x-ray
  absorption spectra (open circles) at the Fe $L_{2,3}$ edges in
  CaCu$_3$Fe$_2$Re$_2$O$_{12}$ measured at 15 K \cite{ShMi14} in comparison
  with the theoretically calculated ones in the GGA+$U$ approximation with
  Hubbard $U$ applied to Cu and Fe sites. The dotted black curve shows the
  background spectrum. Lower panel: the theoretically calculated and experimental XMCD spectra
  of CaCu$_3$Fe$_2$Re$_2$O$_{12}$ at the Fe $L_{2,3}$ edges. The insert at the
  lower panel presents the theoretical Fe $L_3$ XMCD spectra
  calculated in the GGA+$U$ approximation with Hubbard $U$ applied only to the Cu
  site (the green curve) and Cu, Fe, and Re sites simultaneously (the black curve) in
  comparison with the experimental spectrum from Ref. \cite{ShMi14}. }
\end{figure}

Figure \ref{xmcd_Fe_CCFRO} (the upper panel) shows the experimental x-ray
absorption spectroscopy spectra (open circles) at the Fe $L_{2,3}$ edges in
CaCu$_3$Fe$_2$Re$_2$O$_{12}$ measured at 15 K \cite{ShMi14} in comparison with
the theoretically calculated ones in the GGA+$U$ approximation with Hubbard
$U$ applied to Cu and Fe sites. The theory reproduces well the energy
position and the shape of the XAS spectra.

The lower panel of Fig. \ref{xmcd_Fe_CCFRO} presents the theoretically
calculated and experimental XMCD spectra of CaCu$_3$Fe$_2$Re$_2$O$_{12}$ at the Fe $L_{2,3}$
edges. The theory reproduces all peculiarities of the experimental spectra,
although it slightly underestimates the intensity of the negative peak at 713
eV at the $L_3$ edge and is not able to reproduce the low energy shoulder at
$\sim$721 eV at the $L_2$ edge. We found that the best agreement between the
theory and experiment is achieved for the GGA+$U$ approximation with Hubbard $U$
applied to Cu and Fe sites. If Hubbard $U$ is applied only to the
Cu site (the green curve in the insert of the lower panel of
Fig. \ref{xmcd_Fe_CCFRO}) or simultaneously to Cu, Fe, and Re sites 
(the black curve in the insert of the lower panel of Fig. \ref{xmcd_Fe_CCFRO}),
there is no satisfactory agreement with the experimental XMCD spectrum.

\begin{figure}[tbp!]
\begin{center}
\includegraphics[width=0.95\columnwidth]{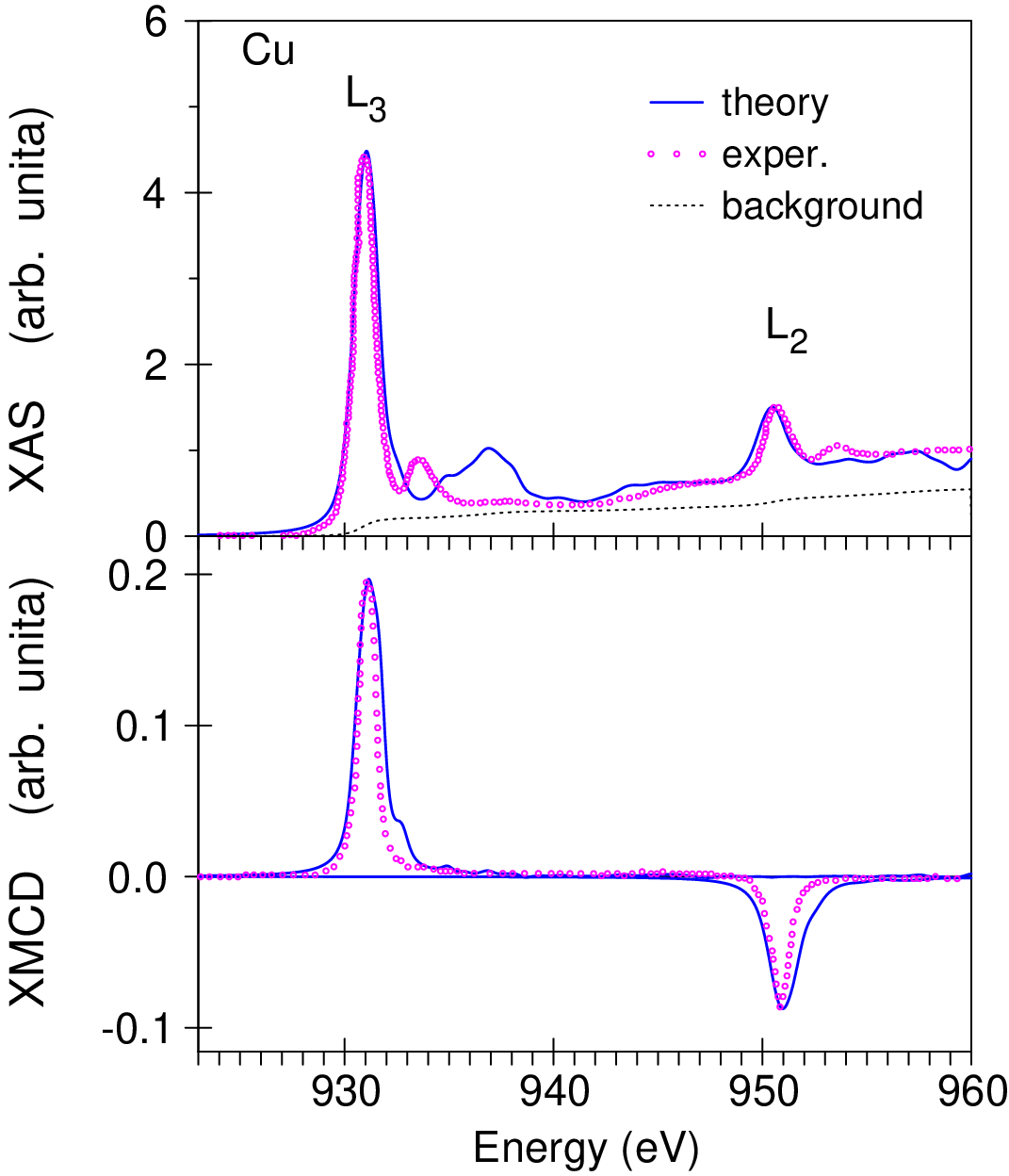}
\end{center}
\caption{\label{xmcd_Cu_CCFRO}(Color online) Upper panel: the experimental XAS
  spectra (open circles) at the Cu $L_{2,3}$ edges in
  CaCu$_3$Fe$_2$Re$_2$O$_{12}$ measured at 15 K \cite{ShMi14} in comparison
  with the theoretically calculated ones in the GGA+$U$ approximation with
  Hubbard $U$ applied to Cu and Fe sites. The dotted black curve shows the
  background spectrum. Lower panel: the theoretically calculated and experimental XMCD spectra
  of CaCu$_3$Fe$_2$Re$_2$O$_{12}$ at the Cu $L_{2,3}$ edges.  }
\end{figure}

Figure \ref{xmcd_Cu_CCFRO} (the upper panel) shows the experimental XAS
spectra (open circles) at the Cu $L_{2,3}$ edges in
CaCu$_3$Fe$_2$Re$_2$O$_{12}$ measured at 15 K \cite{ShMi14} in comparison with
the theoretically calculated ones in the GGA+$U$ approximation with Hubbard
$U$ applied to Cu and Fe sites. The theory reproduces well the energy
position and the shape of the major peak at the $L_3$ edge at 931 eV but the high
energy shoulder at $\sim$933.5 eV is shifted towards higher energy in the
theoretical spectrum. The theory also reproduces well the shape of the $L_3$
XMCD spectrum (the lower panel of Fig. \ref{xmcd_Cu_CCFRO}) except it gives
a small high energy shoulder at 932.8 eV, which is absent in the
experiment. The theoretical $L_2$ XMCD spectrum is slightly wider in
comparison with the experiment.

\begin{figure}[tbp!]
\begin{center}
\includegraphics[width=0.95\columnwidth]{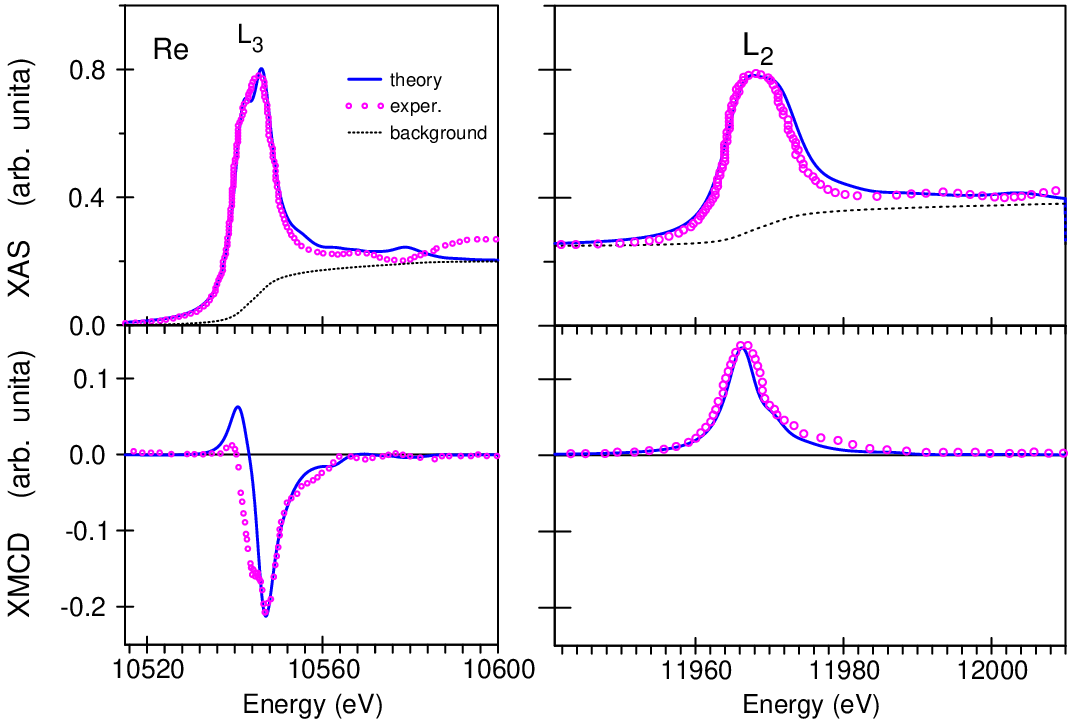}
\end{center}
\caption{\label{xmcd_Re_CCFRO}(Color online) Upper panel: the experimental XAS
  spectra (open circles) at the Re $L_{2,3}$ edges in
  CaCu$_3$Fe$_2$Re$_2$O$_{12}$ measured at 15 K \cite{ShMi14} in comparison
  with the theoretically calculated ones. The dotted black curve shows the
  background spectrum. Lower panel: the theoretically calculated XMCD spectra of
  CaCu$_3$Fe$_2$Re$_2$O$_{12}$ at the Re $L_{2,3}$ edges in comparison with
  the experimental spectra \cite{OBP+17}.  }
\end{figure}

The Re $L_3$ x-ray absorption spectrum possesses a two-peak fine structure
at 10540$-$10548 eV (the upper panel of Fig. \ref{xmcd_Re_CCFRO}), which is
reproduced well in our band structure calculations. The lower energy peak is due to
transitions from Re 2$p_{3/2}$ to Re {\tg} states situated just above the
Fermi level up to 2.2 eV (see Fig. \ref{PDOS_CCFRO}). The second high energy
peak is due to transitions into Re {\eg} states situated at 4.2$-$6 eV above the Fermi level. 
The XMCD signal from the latter transitions is negative, while the signal from the transitions into {\tg}
states is positive and smaller in absolute value. This positive peak is slightly
overestimated in our calculations. Besides, the calculations do not reproduce the low energy shoulder 
of the negative experimental XMCD peak. The agreement between theoretically calculated and
experimentally measured XAS and XMCD spectra at the Re $L_2$ edge is very
good.

\begin{figure}[tbp!]
\begin{center}
\includegraphics[width=0.95\columnwidth]{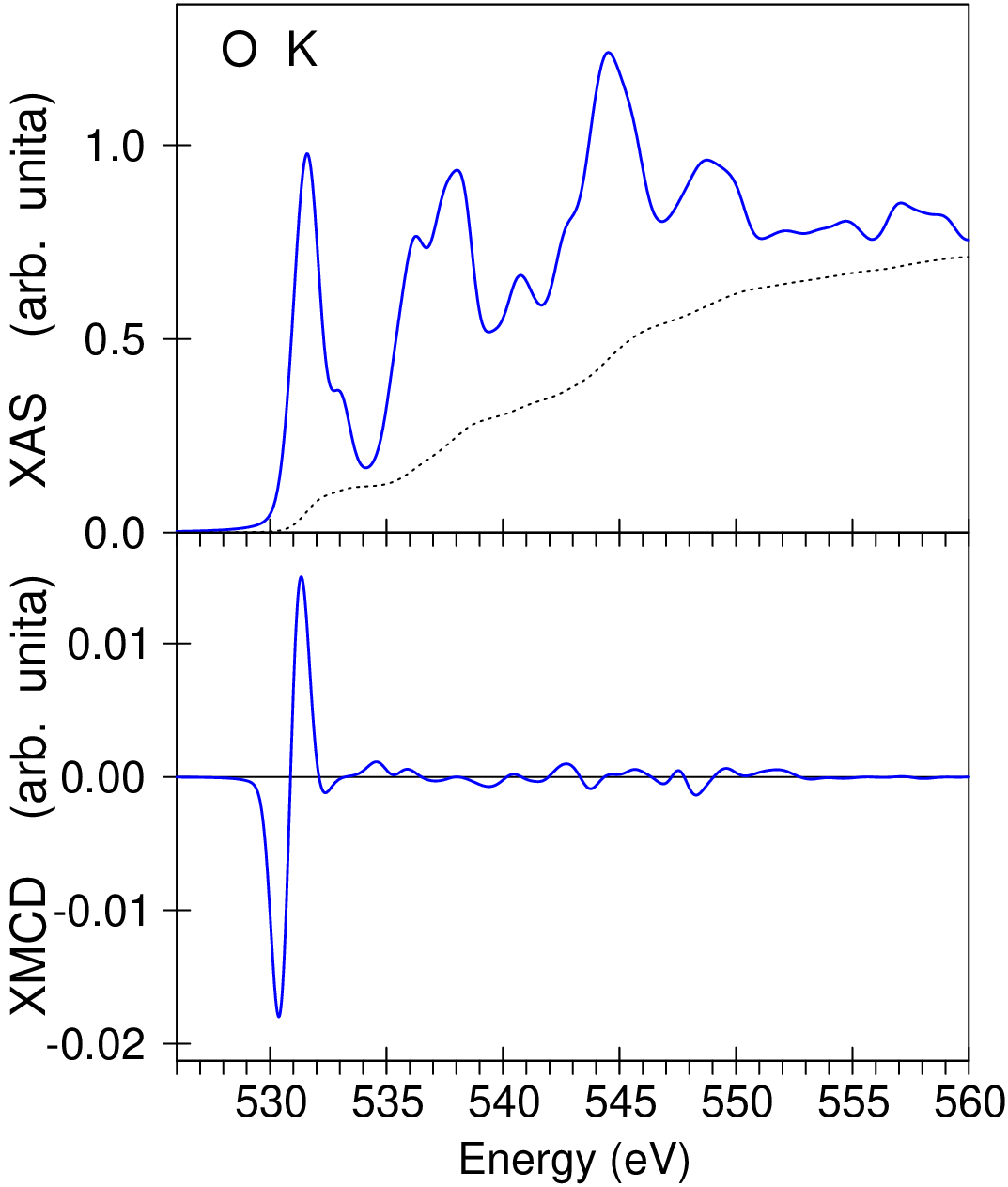}
\end{center}
\caption{\label{xmcd_O_CCFRO}(Color online) Upper panel: the theoretical
  XAS spectrum at the O $K$ edge in CaCu$_3$Fe$_2$Re$_2$O$_{12}$
  calculated in the GGA+$U$ approximation with Hubbard $U$ applied to Cu
  and Fe sites. The dotted black curve shows the background spectrum. Lower
  panel: the theoretically calculated XMCD spectrum of CaCu$_3$Fe$_2$Re$_2$O$_{12}$ at
  the O $K$ edge. }
\end{figure}

Figure \ref{xmcd_O_CCFRO} presents the theoretical XAS (the upper
panel) and XMCD (the lower panel) spectra at the oxygen $K$ edge in
CaCu$_3$Fe$_2$Re$_2$O$_{12}$ calculated in the GGA+$U$ approximation with
Hubbard $U$ applied to Cu and Fe sites. The O $K$ XAS spectrum
possesses a complex structure with many peaks typical to such spectra
\cite{book:AHY04}. The dichroism at the O $K$ edge is much smaller than
the corresponding dichroism at the $L_{2,3}$ edges of Cu, Fe, and Re. 
It is nonzero in a very small energy interval of
$\sim$2 eV just above the $K$ edge where oxygen 2$p$ states are strongly
hybridized with the transition metals $d$ states (see
Fig. \ref{PDOS_CCFRO}). Experimental measurements of the XAS and XMCD
spectra at the oxygen $K$ edge in CaCu$_3$Fe$_2$Re$_2$O$_{12}$ are highly desirable.

\subsection{Magnetic moments}

In magnets, the spin $M_s$ and orbital $M_l$ magnetic moments are basic
quantities and their separate determination is therefore important. Methods of
their experimental determination include traditional gyromagnetic ratio
measurements \cite{Sco62}, magnetic form factor measurements using neutron
scattering \cite{book:MarLov71}, and magnetic x-ray scattering \cite{Blu85}.
In addition to these, the recently developed x-ray magnetic circular dichroism
combined with the sum rules \cite{TCS+92,CTA+93} has attracted much
attention as a method of site- and symmetry-selective determination of M$_s$
and M$_l$. 

\begin{table}[tbp!]
\caption{\label{mom_CCFRO} The theoretical spin $M_s$,
  orbital $M_l$, and total magnetic moments (in {\mb}) calculated for CaCu$_3$Fe$_2$Re$_2$O$_{12}$
  in the GGA and GGA+$U$ approximations. The magnetic moments obtained with the sum rules (\ref{l_z}) 
  and (\ref{s_z}) from the theoretically calculated XAS and XMCD spectra are also presented as well as 
  the magnetic moments estimated in \cite{ShMi14} with the sum rules applied to the experimental spectra. }
\begin{center}
\begin{tabular}{cccccccc}
\hline
method  & atom & $M_s$  & $M_l$ & $M_{total}$ \\
\hline
            & Ca &   0.0128  &  0.0040  &  0.0168   \\
            & Cu &   0.0887  &  0.0031  &  0.0918   \\
GGA         & Fe &   3.8218  &  0.0550  &  3.8768   \\
            & Re &  -0.2114  &  0.0554  & -0.1560   \\
            & O  &   0.0994  &  0.0009  &  0.1003   \\
\hline
            & Ca &  -0.0028  &  0.0086  &  0.0058   \\
            & Cu &   0.2669  &  0.0231  &  0.2901   \\
GGA+U       & Fe &   4.0703  &  0.0615  &  4.1318   \\
            & Re &  -0.6358  &  0.1586  & -0.4773   \\
            & O  &   0.1129  &  0.0019  &  0.1147   \\
\hline
GGA+U       & Cu &   0.4408  &  0.0383  &  0.4791  \\
sum         & Fe &   4.1395  &  0.0626  &  4.2021  \\
rules       & Re &  -0.5698  &  0.1309  & -0.4389   \\
\hline
exper.       & Cu&   0.82    &   0.04  &  0.86  \\
\cite{ShMi14}& Fe&   4.11    &   0.06  &  4.17  \\
\hline
\end{tabular}
\end{center}
\end{table}

Table \ref{mom_CCFRO} presents the theoretically calculated
in the GGA and GGA+$U$ approximations spin $M_s$, orbital $M_l$, 
and total magnetic moments in CaCu$_3$Fe$_2$Re$_2$O$_{12}$.

The Cu and Fe spin and orbital moments are parallel, whereas the spin and
orbital Re moments are antiparallel, which is in accordance with Hund's third rule. 
Our calculated GGA orbital magnetic moments
for Cu, Fe, and Re are lower than the results of the GGA+$U$ approach. 
It should be mentioned that the effect of the Coulomb
correlations changes the energy band structure of transition metal compounds in
two ways. First, the occupied $d$ states are shifted downward by $U_{\rm{eff}}$/2 and
the empty $d$ states are shifted upwards by this amount relative to the Fermi
energy. Second, the Coulomb correlations enhance an effective spin-orbit
coupling constant \cite{LAJ+08}. 

We present in Table \ref{mom_CCFRO} the results obtained with the sum
rules (\ref{l_z}) and (\ref{s_z}) applied to the theoretically calculated XAS
and XMCD spectra. We found that the sum rules reproduce the GGA+$U$ values of
the spin magnetic moment within 39\%, 2\%, and 10\% for Cu, Fe, and Re,
respectively. The orbital magnetic moments are reproduced within 39\%, 2\%,
and 17\% for Cu, Fe, and Re, respectively.

Table \ref{mom_CCFRO} also shows the magnetic moments estimated in Ref. \cite{ShMi14} 
for Cu and Fe by applying the sum rules to the experimentally measured XAS and XMCD spectra. 
There is quit good agreement between the calculated GGA+$U$ values of the Fe
spin and orbital moments (4.0703 and 0.0615 {\mb}, respectively), and the
values obtained by applying the sum rules to the theoretical (4.1395 and 0.0626
{\mb}) and experimental (4.11 and 0.06 {\mb}) XAS and XMCD spectra. However,
the corresponding agreement for the Cu magnetic moments is somewhat worser.

Finally, our calculations produce induced spin and orbital magnetic moments at
the oxygen site in CaCu$_3$Fe$_2$Re$_2$O$_{12}$ of about 0.0994 {\mb} and
0.0009 {\mb}, respectively, in the GGA approach and 0.1129 {\mb} and 0.0019
{\mb} in GGA+$U$. The calculated 5$d$ spin and orbital
magnetic moments at the Ca site were found to be equal to $M_s$ = $-$0.0028
{\mb} and $M_l$ = 0.0086 {\mb} in the GGA+$U$ approach.

\section{Summary}
\label{sec:summ}

The electronic and magnetic structures as well as the XAS and XMCD spectra
of the A- and B-site ordered perovskite CaCu$_3$Fe$_2$Re$_2$O$_{12}$ 
have been investigated theoretically within the
GGA and GGA+$U$ approximations in the framework of the fully relativistic
spin-polarized Dirac LMTO band-structure method.

The electron-electron correlations play an important role in the electronic
structure and physical properties of CaCu$_3$Fe$_2$Re$_2$O$_{12}$. Hubbard
$U$ shifts the occupied and empty states by $U_{\rm{eff}}$/2 downwards and
upwards, respectively. As a result, energy gaps appear in the spin-up
states at the Cu, Fe, Re, and oxygen sites. Besides, the Cu and Fe 3$d$ states spread
in a wider energy interval and change the shape. The Coulomb correlations also
enhance an effective spin-orbit coupling constant. It leads to increasing of the
orbital magnetic moments at the transition metal sites. The calculated Cu and Fe spin and
orbital moments are parallel, whereas the spin and orbital Re moments are
antiparallel in CaCu$_3$Fe$_2$Re$_2$O$_{12}$, which is in accordance with Hund's third
rule. Although Hubbard $U$ was not applied to oxygen states, they are also
changed due to the strong hybridization with the Fe 3$d$ and Re 5$d$ states.

We have studied the XAS and XMCD spectra at the Cu, Fe, Re $L_{2,3}$ and oxygen $K$ edges. 
We found that the best agreement between the theory and experiment is achieved for the GGA+$U$ 
approximation with Hubbard $U$ applied to Cu and Fe sites. If Hubbard $U$ is applied only to the
Cu site or simultaneously to Cu, Fe, and Re sites, there is no satisfactory agreement 
with the experimental XMCD spectrum at the $L_3$ edge.
The calculations show good agreement with the experimental measurements of the
XAS and XMCD spectra at the transition metal $L_{2,3}$ edges. The oxygen $K$
XAS spectrum possesses a complex structure with many peaks typical to such spectra. The dichroism 
at the oxygen $K$ edge is much smaller than the corresponding dichroism at the $L_{2,3}$ edges of Cu, Fe, and Re. 
It is nonzero in a very small energy interval of $\sim$2 eV just above the $K$ edge where oxygen 2$p$ states 
are strongly hybridized with the transition metals $d$ states.

There is quit good agreement between the calculated GGA+$U$ Fe spin and orbital
magnetic moments and the values obtained by applying the sum rules to the theoretical and
experimental XAS and XMCD spectra. The corresponding agreement for the Cu magnetic moments is
somewhat worser.


\newcommand{\noopsort}[1]{} \newcommand{\printfirst}[2]{#1}
  \newcommand{\singleletter}[1]{#1} \newcommand{\switchargs}[2]{#2#1}

\end{document}